\documentclass[12pt]{article}
\usepackage{amsfonts,amsmath}

\usepackage{color}
\usepackage{tabularx}

\usepackage{dsfont}
\usepackage{times}

\usepackage[dvips]{graphicx}
\usepackage{ifthen}
\usepackage{enumerate}
\usepackage{amsmath}
\usepackage{amssymb}
\usepackage[mathscr]{eucal}
\usepackage[errorshow]{tracefnt}
\usepackage{amsmath}
\usepackage{slashed}
\usepackage{amssymb}
\usepackage{array}
\usepackage{amsthm}
\usepackage{eucal}
\usepackage{eufrak}
\usepackage{epsf}
\usepackage{epsfig}
\usepackage{psfrag}
\usepackage{multirow}
\usepackage[apple mac]{inputenc}

\newcommand{\complex}{{\mathds C}} 

\def\+{{+\!\!\!+}} 
\def\pp{\mbox{\tiny${}_{\stackrel\+ =}$}}

\def\complex      {{\mathbb C}}

\def\pp   {{\mathbb P}}

\def\reals        {{\mathbb R}}
\def\zet          {{\mathbb Z}}

\def\pmb#1{\setbox0=\hbox{#1}%
\kern.0em\copy0\kern-\wd0 
\kern-.04em\copy0\kern-\wd0 
\kern.08em\copy0\kern-\wd0 
\kern-.04em\raise.0433em\box0 }         
 
\newcommand{\nc}{\newcommand} 
\nc{\beq}{\begin{equation}} 
\nc{\eeq}[1]{\label{#1}\end{equation}} 
\nc{\ber}{\begin{eqnarray}} 
\nc{\eer}[1]{\label{#1}\end{eqnarray}} 
\nc{\pek}[1]{\cite{#1}} 
\nc{\enr}[1]{(\ref{#1})} 
\nc{\kal}[1]{{\cal{#1}}} 
\nc{\dott}{\;\cdot\;} 

\def\0 {\nonumber}

\begin{document} 

\setcounter{page}{0}
\newcommand{\inv}[1]{{#1}^{-1}} 
\renewcommand{\theequation}{\thesection.\arabic{equation}} 
\newcommand{\be}{\begin{equation}} 
\newcommand{\ee}{\end{equation}} 
\newcommand{\bea}{\begin{eqnarray}} 
\newcommand{\eea}{\end{eqnarray}} 
\newcommand{\re}[1]{(\ref{#1})} 
\newcommand{\qv}{\quad ,} 
\newcommand{\qp}{\quad .} 

\def\qp{Q_+}
\def\qm{Q_-}
\def\qbp{\bar Q_+}
\def\qbm{\bar Q_-}
\def\sgh{\Sigma_{g,h}}

\begin{titlepage} 
\begin{center} 

\hfill SISSA 60/2009/EP-FM\\  
                         
\vskip .3in \noindent 


{\Large \bf{Hitchin systems, ${\cal N}=2$ gauge theories and W--gravity}} \\

\vskip .2in 

{\bf Giulio Bonelli and Alessandro Tanzini}

\vskip .05in 
{\em\small International School of Advanced Studies (SISSA) \\ and \\ INFN, Sezione di Trieste \\
 via Beirut 2-4, 34014 Trieste, Italy} 
\vskip .5in
\end{center} 
\begin{center} {\bf ABSTRACT }  
\end{center} 
\begin{quotation}\noindent  

We propose some arguments supporting an M-theory derivation of the duality recently discovered by Alday, Gaiotto and Tachikawa
between two-dimensional conformal field theories and ${\cal N}=2$ superconformal gauge theories
in four dimensions. We find that $A_{N-1}$ Toda field theory is the simplest 
two-dimensional conformal field theory quantizing the moduli of $N$ M5-branes wrapped on a Riemann surface.
This leads us to identify chiral operators of the ${\cal N}=2$ gauge theories with integrated W-algebra currents.
As a check of this correspondence we study some relevant OPE's obtaining that Nekrasov's partition function satisfies
W-geometry constraints.
\end{quotation} 
\vfill 
\eject 

\end{titlepage}


\section{Introduction} 

M5-branes can -- and assuming string dualities, have to -- be used to engineer gauge theories in lower dimensions via 
compactifications.
Because of the absence of coupling constants in M-theory, the gauge theories which one obtains in this way have 
a space of couplings which is completely geometrical in nature. This makes these theories quite appealing.
On the other hand, they are generically in a strongly coupled phase. Actually, since one starts from an unknown theory, that of the 
M5-branes, one gets generically non-Lagrangian theories. 
These can be specified only once the geometry of the compactification is realized to correspond to a perturbative corner
in the lower dimensional gauge theory.
A way to fix the dictionary is to link the perturbative descriptions of these corners in a way which is consistent with the geometry at hand.
This program, which has been initiated in \cite{K,four}, led to the formulation of a new class of ${\cal N}=2$ 
theories in \cite{N=2} upon reduction on a Riemann surface.

This approach stimulated further studies which resulted in a new exciting realization of dualities between two and four dimensional 
conformal field theories \cite{AGT}. This consists in a CFT$_2$ interpretation of the $SU(2)$ Nekrasov partition function \cite{N} in terms of 
the conformal blocks of Liouville gravity. This has been extended to higher rank theories in \cite{W}, the CFT$_2$ in this 
case being proposed to be the Toda field theory. Evidence of this duality and its consequences have been presented in \cite{tutti}.
A derivation in the context of a B-model topological string theory setup has been proposed in \cite{DV}.

In this note we propose an M-theory derivation
of the duality. We consider the ${\cal N}=2$ SCFT obtained by wrapping $N$ M5-branes on a Riemann surface 
$\Sigma$ and derive a corresponding Toda theory on $\Sigma$ itself describing its moduli. 
We focus on the $A_{N-1}$ case, but our construction should extend to an arbitrary simply laced Lie algebra of $ADE$ type.
We remark that in the study of M-theory compactifications, Hitchin systems play a central r\^ole \cite{DW,GMN}; this is
our starting point.
In particular, we show that the Hitchin system associated to the wrapped $N$ M5-branes can be described as the Toda field theory
which we recognize to be the natural CFT$_2$ quantizing the geometric moduli of the problem.
Moreover, we identify the chiral ring of the ${\cal N}=2$ gauge theory in terms of the $W_N$-currents of the Toda system and we provide
a check of this correspondence by studying the relevant OPE algebra.

\section{From M5-branes to Toda field theory}

Let us start by considering the geometry of a system of $N$ M5-branes wrapped on a manifold $Y_6$ 
with topology $\Sigma\times\reals^4\times \{pt\}$ 
in $T^*\Sigma\times\reals^4\times\reals^3$ where $\Sigma$ is a Riemann surface. This can be equipped with a non trivial
fibration of $\reals^4$ over $\Sigma$ which specifies the $\Omega$-background of Nekrasov \cite{NO}. 
The geometry of the M5-branes bound state is described by an $N$-fold branched covering of $\Sigma$ given
by the algebraic equation
\be
x^N=\sum_{j=2}^N W_j(z) x^{N-j}
\label{1}
\ee
where $x$ is a section in the fiber in $T^*\Sigma$ and $W_j$ are $(j,0)$-holomorphic differentials on $\Sigma$.

We want to describe the simplest, that is the minimal, conformal field theory quantizing the moduli of the problem. 
This can be achieved by realizing the geometric curve (\ref{1}) as the spectral curve of an Hitchin system
\cite{NH} whose flat $SL(N,\complex)$ connection reads
\be
\nabla = \left(\partial + X\right) + \bar\partial
\label{DS}\ee
where 
\be
X= \begin{pmatrix}
  0 & 1       &     &        &    &   \\
    & 0       &  1  &        &    &   \\
    &         &     & \vdots &    &   \\
    &         &     &        & 0  & 1 \\
W_N & W_{N-1} &     & \ldots & W_2& 0
\end{pmatrix}
\label{ok}
\ee
We will refer to this gauge as the Drinfel'd-Sokolov one.

In order to define a unitary field theory quantizing the above moduli, we have to bring the above connection
to an  $SU(N)$ form. This can be achieved \cite{AF} by a conjugation via an $SL(N,\complex)$ element parametrized by a Cartan 
valued real field $\Phi$. The Hitchin equations describing the BPS bound state are 
\bea
F_A+[\beta,\beta^\dagger]=0
\\
\bar\partial_A\beta=0
\label{hitchin}\eea

Choosing 
\bea
\beta={\cal E}_+ dz \label{questa}
\\
A_{\bar z}=0 \nonumber
\eea
with $A$ the metric connection associated to $H=exp\Phi$, $\Phi\in sl(N,\reals)$ a real field taking values in the Cartan subalgebra, 
it follows that
\bea
\beta^\dagger= e^{-\Phi}{\cal E}_-e^\Phi d\bar z \label{quella}
\\
A_z=\partial_z\Phi \nonumber
\eea
In (\ref{questa}) and (\ref{quella}), ${\cal E}_\pm$ is the sum of the $A_{N-1}$ positive(negative) simple roots.
The Hitchin equations are equivalent to the flatness of the spectral connection 
\be
\nabla=\partial_A + \beta + \left(\partial_A + \beta\right)^\dagger= \partial + {\cal A} + \left(\partial + {\cal A}\right)^\dagger.
\label{spectre}\ee
Under the gauge transformation $g=e^{-\Phi/2}$ the spectral connection gets manifestly unitary and reads
\bea
{\cal A}_z^g=\frac{1}{2}\partial_z\Phi + \exp\left({\frac{1}{2}ad_\Phi}\right){\cal E}_+ 
\nonumber\\
{\cal A}_{\bar z}^g=-\frac{1}{2}\partial_{\bar z}\Phi + \exp\left({-\frac{1}{2}ad_\Phi}\right) {\cal E}_-
\label{greg}\eea

The flatness condition for ${\cal A}^g$ in (\ref{greg}) are the Toda $A_{N-1}$ field equations on $\Sigma$
\be
\partial_z\partial_{\bar z}\Phi=\sum_i h_i e^{\alpha_i(\Phi)}
\ee
where $h_i$ are the Cartan elements of the Lie algebra and $\alpha_i$ its simple roots
\footnote{Notice that in the above the Toda parameter is $b=1$. As far as the classical theory is concerned, a generic value of $b$ 
can be achieved 
via a field rescaling. At quantum level one has to consider it as a background charge parametrizing the non trivial 
fibration of the $\Omega$-background as in \cite{DV}.}.
On the other hand one can prove (see theorem 4.1 of \cite{AF} and \cite{AF2}) that there exists a gauge transformation which brings 
the Toda connection
(\ref{spectre}) in the Drinfel'd-Sokolov form (\ref{DS}).

The holomorphic differentials $W_i$'s are realized then as polynomials in the holomorphic derivatives 
of the elements of the field $\Phi$ via the Miura transform 
$$
{\rm det}(\partial-\partial\Phi)=\partial^N-\sum_{j=0}^{N-2}W_{N-j}\partial^j
$$ 
and their conservation $\partial_{\bar z} W_i=0$ follows from the above flatness condition\footnote{See for example \cite{BO}.}.


Let us show the above construction in detail for the Liouville case $N=2$.
We have 
\be
{\cal A}_z=\begin{pmatrix}
\partial_z\varphi & 1 \\
0 & -\partial_z\varphi
\end{pmatrix}
\,\,\,\, , \quad
{\cal A}_{\bar z}=\begin{pmatrix}
0 & 0 \\
e^{2\varphi} & 0
\end{pmatrix}
\label{liouville}
\ee
whose flatness condition is the Liouville equation $\partial_z\partial_{\bar z}\varphi=e^{2\varphi}$.
The gauge transformation \\
$
\gamma=\tiny
\begin{pmatrix}
  1 & 0 \\
-\partial_z\varphi &1
\end{pmatrix}
$
leads to
\bea \nonumber
{\cal A}^\gamma_z&=&\gamma^{-1}\left({\cal A}_z + \partial_z\right)\gamma=
\begin{pmatrix}
  0 & 1 \\
  \left(\partial_z\varphi\right)^2-\partial_z^2\varphi  &0
\end{pmatrix}
\\ 
{\cal A}^\gamma_{\bar z}&=&\gamma^{-1}\left({\cal A}_{\bar z} + \partial_{\bar z}\right)\gamma=
\begin{pmatrix}
  0 & 0 \\
  e^{2\varphi}-\partial_z\partial_{\bar z}\varphi  & 0
\end{pmatrix}
\eea
which reproduces (\ref{ok}) with $W_2=\left(\partial_z\varphi\right)^2-\partial_z^2\varphi$
because of the flatness condition. Actually, the Liouville equation implies both the holomorphicity of the Drinfel'd-Sokolov 
connection and its flatness $\bar\partial W_2=0$.
Finally, the specific form of (\ref{greg}) for $N=2$ is
\be
{\cal A}^g_z=
\begin{pmatrix}
  \frac{1}{2}\partial_z\varphi  & e^\varphi \\
  0 & -\frac{1}{2}\partial_z\varphi 
\end{pmatrix}
\,\,\, ,\quad
{\cal A}^g_{\bar z}=
\begin{pmatrix}
  -\frac{1}{2}\partial_{\bar z}\varphi  & 0 \\
  e^\varphi & \frac{1}{2}\partial_{\bar z}\varphi 
\end{pmatrix}
\ee
and the Miura transform reads $\det\left(\partial-\partial\Phi\right)=\left(\partial-\partial\varphi\right)\left(\partial+\partial\varphi\right)
=\partial^2-(\partial\varphi)^2+\partial^2\varphi$.


We therefore conclude that the simplest unitary field theory for the BPS moduli of the problem above is the $A_{N-1}$ Toda theory.
It is indeed the case that W-gravity is the right framework to quantize the space of higher differentials on $\Sigma$
generalizing the usual 2D gravity \cite{bilal}.
Let us remark that one could have expected this result by requiring the theory to arise as 
an interacting deformation of the free bosonic fields via a potential preserving both conformal 
invariance and permutation symmetry. 

In the following subsections we collect some consistency checks and some consequences of the above observations.

\subsection{${\cal N}=2$ chiral ring and W-algebrae}

In this subsection we discuss the $CFT_2$ dual of BPS local operators of the four dimensional ${\cal N}=2$ gauge theory.

Let us notice that the six dimensional $SO(6)$ Euclidean group is naturally broken to $U(1)\times SO(4)$ by the 
fibered structure of the M5-branes world-volume. 
The U(1) factor is seen from the four dimensional point of view as R-symmetry, while it is the rotational 
symmetry in two dimensions. We therefore conclude that the R-symmetry charge in the ${\cal N}=2$ gauge theory has 
to be identified with the conformal spin in two dimensions.

Let us consider now a corner of the moduli space of the $A_{N-1}$-theory where a long tube is displayed on $\Sigma$.
According to \cite{N=2}, in such a situation an $SU(N)$ vector multiplet becomes weakly coupled.
It is therefore meaningful to consider local operators composed by its scalar component $\phi$, namely $Tr(\phi^j)(P)$ with 
$j=2,\ldots,N$, where $P$ is a point in the four dimensional fiber.

Consider the case $j=2$ first.
The insertion of this operator, in the presence of the $\Omega$-background, has been computed in \cite{LMN,FMP,ale}
and shown to correspond to the derivative of the Nekrasov partition function with respect to the 
corresponding complex gauge coupling.
According to the mapping of the gauge coupling to the moduli space of $\Sigma$, this corresponds in the two dimensional dual picture
to a derivative with respect to the associated geometric modulus. This is realized in the $CFT_2$ by the insertion of the integrated operator
given by the stress energy tensor $T=W_2$ times the Beltrami differential $\mu_{collar}^{(-1,1)}$, 
that is $\int_\Sigma \mu_{collar}^{(-1,1)} T$, 
where $\mu_{collar}^{(-1,1)}$ is the Beltrami differential associated to the modulus of the long collar where it has support.
These have been studied for example in \cite{Wolpert}. 
Locally, where the collar is the annulus $A_\tau=\{|\tau|<|z|<1\}$, the Beltrami differential is given by $\partial_{\bar z}v^z$  
where $v^z=\frac{z\ln |z|}{\tau\ln |\tau|}$ is the vector field associated to the infinitesimal stretch in $\tau$. The above Beltrami 
differential is of course dual to the quadratic holomorphic differential $\frac{-\tau(dz)^2}{\pi z^2}$ and $\tau$ is identified with the 
Coulomb branch parameter $u_2=\frac{1}{2}\langle Tr\phi^2\rangle$.
This corresponds, from the M-theory point of view, to an operator generated by an M2-brane ending on $\Sigma \times P$ supported on the stretching collar.

The generalization to higher $j$ comes natural, the statement being that similarly the insertion of the higher $j$ BPS operators  
above corresponds to the insertion of the integrated j-current
\be
\langle Tr(\phi^j)\ldots\rangle_{{\cal N}=2}\sim \langle \int \mu_{collar}^{(-j+1,1)} W_j \ldots\rangle_{CFT_2}
\label{dictionary}\ee
 where $W_j$ are the higher differentials
in the Toda CFT and $\mu_{collar}^{(-j+1,1)}=\partial_{\bar z}\left(\frac{z\ln |z|}{\ln |\tau|}\right)^{j-1}$ 
the corresponding generalized Beltrami differentials supported on the stretching collar.
Indeed, by generalizing the computation in \cite{Wolpert}, one can show that $\mu_{collar}^{(-j+1,1)}$ is dual to the 
$j$-differential $\frac{(dz)^j}{z^j}$ and therefore its overlap with $W_j$
extracts exactly its maximal residue. More explicitly, if $W_j=\frac{u_j (dz)^j}{\pi z^j}+\ldots$, then $\int_{A_\tau}
\mu_{collar}^{(-j+1,1)}W_j=u_j$.

This point of view implies that the space of protected couplings for these ${\cal N}=2$ $A_{N-1}$ theories coincides with 
the $W_N$ Hitchin extended moduli space of the Riemann surface.
The assignment of punctures in the general case is more subtle than for $j=2$.
Indeed, normalized quadratic differentials at the punctures are specified by a pure quadratic pole. Instead
normalized $j$-differentials at the punctures are specified by a pole of order $j$.
For $n$ punctures of general type and for negative Euler characteristics, the space of holomorphic $j$-differentials has 
complex dimension $d_j=(2j-1)(g-1)+(j-1)n$.
However, one can further constrain lower polar parts for higher differentials at the punctures and therefore 
change their geometric nature. This clearly changes also the dimension 
according to
$d_j=(2j-1)(g-1)+\sum_{t=1}^{j}(j-t)n_{j,t}$
where $t$ denotes the type of the puncture, starting from $t=1$ which denotes general punctures.
The total number of punctures is given by $n=\sum_t n_{j,t}$. 
If all $j$-differentials have no polar structure at a given point, that results just in a formal puncture
and should not be counted.

In the cases studied in \cite{N=2} the $j$-differentials are singular at all the punctures, namely $n_{j,j}=0$ for all $j$'s.
For example, in the $SU(3)$ case one can limit to two kinds of punctures only. On a Riemann surface with 
genus $g$ and $n_1$ punctures of general type and $n_2$ constrained punctures 
one has $n_{2,1}=n_1+n_2$, $n_{3,1}=n_1$ and $n_{3,2}=n_2$, so that
$d_2=3(g-1)+n_1+n_2$ and $d_3=5(g-1)+2n_1+n_2$,
which matches the field theory counting in \cite{N=2}.
We notice that in order to produce massive deformations for the superconformal gauge theories one should allow
simple poles in the SW differential $d\lambda =  xdz$ and thus double poles for the quadratic differentials.

A more stringent check of the correspondence is provided by the comparison of the $W$-algebra OPE and the appropriate equation satisfied by 
local BPS operator insertions in the gauge theory.
Actually, as it was shown in \cite{ale}, the insertion of the BPS local operators $Tr\phi^j$
satisfies the equation
\be
\langle \frac{1}{2} Tr\phi^2 Tr\phi^j\rangle - \langle \frac{1}{2} Tr\phi^2\rangle \langle Tr\phi^j\rangle 
= - q\partial_q \langle Tr\phi^j\rangle
\label{uno}
\ee
This corresponds in the Toda field theory to the OPE\footnote{After a redefinition of the higher $W_j$'s, see \cite{bilal}.} 
\be
W_2(z)W_j(w)=\left(\frac{j}{(z-w)^2}+\frac{\partial_w}{z-w}\right)W_j(w)
\label{due}
\ee
integrated over the appropriate generalized Beltrami differentials on the stretching collar.
Indeed, for any Conformal Field Theory on a Riemann surface $\Sigma$ the very definition of the
stress energy tensor is via the Teichmuller deformation of the amplitude of a generic conformal 
invariant observable $X$
\be
\delta_{Teich}\left(\langle X \rangle Z\right)= Z \sum_a \delta m_a \int_\Sigma \mu_a^{(-1,1)}(z)
\langle T(z) X \rangle d^2z
\label{teich}
\ee
where $Z$ is the CFT partition function, $\mu_a$ is a basis of Beltrami differentials and $m_a$ the associated moduli coordinates
(see for example eq (11) of \cite{EO}).
Notice that eq.(\ref{teich}) is usually rewritten in terms of connected correlation functions as
\be
\delta_{Teich}\langle X \rangle = \sum_a \delta m_a \int_\Sigma \mu_a^{(-1,1)}(z) \langle T(z) X\rangle_{conn}
\label{teichconn}\ee
Since we want to compare with instanton dominated correlation functions, we need to consider a corresponding weakly 
coupled gauge sector. To obtain this we need to consider a Riemann surface with a stretching collar.
Thus we focus on the relevant Beltrami differential $\mu_{collar}^{(-1,1)}$ associated to this degeneration of the Riemann surface.
The modulus of the stretching collar $m_{collar}=\tau$ with large imaginary part is related to the inverse gauge coupling constant, so that 
$q=e^{2\pi i \tau}$ weights perturbatively the instanton contributions to the partition function.

Let us check the correspondence between higher $j$ BPS operators and integrated $W_j$ currents from eq.(\ref{teich}) applied 
to $X=\int_\Sigma \mu_{collar}^{(-j+1,1)}W_j$ by selecting the relevant Beltrami differential $\mu_{collar}^{(-1,1)}$.
Indeed the OPE (\ref{due}) implies that $X$ has the suitable conformal weight.
Thus eq.(\ref{teichconn}) gives
\be
-q\partial_q  \langle \int_\Sigma \mu_{collar}^{(-j+1,1)}W_j \rangle= 
\langle \left( \int \mu_{collar}^{(-1,1)}W_2\right)\left( \int \mu_{collar}^{(-j+1,1)}W_j\right) \rangle_{conn}
\ee 
which is equation (\ref{uno}) under the BPS operators/CFT integrated currents correspondence proposed above.

Developing the full picture therefore requires a generalization of (\ref{teichconn}) to the Hitchin moduli space, namely for any $j$
\be
\delta_{Hitchin}\langle X \rangle = \sum_a \delta m_a \int_\Sigma \mu_a^{(-j+1,1)}(z) \langle W_j(z) X\rangle_{conn}
\label{hitchinconn}\ee
as in \cite{bilal}.
The gauge theory counterpart of this amounts to 
consider an extended Nekrasov partition function 
$Z_{ext}=\langle e^{\sum_j \tau_j Tr\left(\phi^j\right)}\rangle$
where we
switched on
the parameters $\tau_j$ of the Coulomb phase deformation.
The full set of W-gravity constraints should then translate in differential equations on the higher Casimir's
in $Z_{ext}$ of which eq.(\ref{uno}) is a particular case.
A detailed study of this issue will be presented elsewhere.

We therefore argue that for $A_{N-1}$ theories on $\Sigma$, the moduli space of BPS deformations 
is given by the $W_N$-extension of the Teichmuller moduli space to higher differentials.
Let us remark that once one recognizes the Toda theory/W-gravity to be the theory quantizing the extended Teichm\"uller space of Hitchin,
one can extend the analysis of \cite{surface} and \cite{drukker} to the higher rank case.
In \cite{surface} it was shown that the Liouville dual of loop operators of the four dimensional gauge theory 
are given by the holonomies of the connection (\ref{ok}) for $N=2$.
It is then natural to extend the duality at higher rank by considering the holonomies of the $A_{N-1}$ Toda connection (\ref{greg}).
As far as the surface operators are concerned, the generalization of the Liouville dual proposed in \cite{surface}
is provided by the null operators of the Toda field theory.

\section{Comments and open questions}

We conclude with some further observations and open questions concerning the gauge/Toda duality.


The classical configurations of the Toda field theory define holomorphic maps
of $\Sigma$ in $\complex \pp^{N-1}$.
We briefly recall here the line of arguments leading to this result and refer to 
\cite{bilal,hunif} for further details.
One can study the linear system associated to the Hitchin spectral connection $\nabla\Psi=0$.
In the DS-gauge (\ref{DS}) this reads
\be
\left(\partial +X\right)\Psi=0
\quad {\rm and} \quad
\bar\partial\Psi=0
\label{als}
\ee
The set of solutions to (\ref{als}) define an holomorphic map
from the Riemann surface $\Sigma$ to $\complex \pp^{N-1}$ which is associated to every solution of the Toda equations.
One can expand (\ref{als}) in components of $\Psi$ reducing it to a $N$-th order linear differential equation 
on the top component. The $N$ solutions of such a system specify the holomorphic map. 
We are therefore mapping classical Toda field configurations to BPS classical configurations of a
$\complex \pp^{N-1}$ $\sigma$-model on $\Sigma$.
This could have some relevance for the study of surface operators. Actually
the push forward of $\Sigma$ via the holomorphic map is an element in 
$H_2\left(\complex \pp^{N-1},\zet\right)$. Notice that this homology group labels
the magnetic charges of surface operators in the gauge theory \cite{surface}. 
On the other hand, counting holomorphic maps points to the A-model and this suggests an interpretation of the Toda conformal blocks 
as the refined topological vertex of \cite{IKV}.
Indeed, we considered the M5-brane bound system on $T^*\Sigma\times\reals^4\times\reals^3$. 
One could try to generalise this approach to other four-manifolds by fibering
them on the Riemann surface and wrapping M5-branes on the resulting total space.

A relevant issue is the derivation of the $CFT_2$ central charge in the context of
$A_{N-1}$ theory.  This 
should come from the conformal anomaly of M5-branes upon reduction on the twisted geometry
giving rise to Nekrasov's $\Omega$-background. The idea is to integrate the anomaly eight-form of the $N$ M5-brane
system over the four-dimensional space-time to get the anomaly four-form in two-dimensions, whose coefficient should give the central charge. 
The anomaly eight-form reads \cite{anomaly}
\be
I[A_{N-1}]=(N-1)\left[I[1]+N(N+1)\frac{p_2({\cal N}_{Y})}{24}\right] ,
\label{mmmm}\ee
where ${\cal N}_{Y}$ is the normal bundle of $Y_6$ and $p_2$ its second Pontryagin.
In (\ref{mmmm}) $I[A_{N-1}]=I[N]-I[1]$, where $I[N]$ is the anomaly of $N$ M5-branes.
We notice that after subtracting the $U(1)$-factor, the $N$ dependence of (\ref{mmmm}) matches the one of the Toda central charge
\be
c=(N-1)\left(1+Q^2 N(N+1)\right)
\ee
A detailed matching would provide a stronger evidence of the completeness of the $CFT_2$ description at the quantum level
\footnote{This expectation has been confirmed, 
short after the submission of this paper to the arXiv, in \cite{ciapa} where 
the computation of the central charge has been performed along the lines we suggested, finding full agreement.}.
The subtraction of the $U(1)$ center-of-mass of the M5-brane system could be related to the decoupling of
a $U(1)$ factor used in \cite{AGT} to match the four and two-dimensional theories.
We observe that (\ref{mmmm}) has been used in \cite{benini} to reproduce the conformal anomaly of four-dimensional
gauge theories; also in that case the $U(1)$ subtraction revealed to be crucial.

It is known \cite{ale} that the correlators with gaugino insertions can be evaluated from those of ${\cal N}=2$ chiral operators
by minimizing the superpotential on the Coulomb branch. It would be interesting to understand the corresponding construction in W-gravity.
This should correspond to study correlators at fixed momenta.

There is a well-known link between CFTs in two dimensions and Chern-Simons theories in three.
In particular for Toda theories this has been analyzed in \cite{bilal,LO}, where the conformal blocks
of the Toda theory were argued to correspond to the $SU(N)$ Chern-Simons wave functionals on $\Sigma\times \reals$.
On the other hand, the worldvolume theory of M2-branes has been formulated in terms of Chern-Simons theories
\cite{membrane}. It is tempting to identify the space $\Sigma\times \reals$ as the world-volume of 
M2-branes ending on the M5-branes and to analyze the canonical quantization of the M2-branes theory 
in this perspective.

A derivation of the Toda/gauge correspondence has been recently proposed in \cite{DV} in the mirror B-model
framework. It would be very nice to link the picture described in this paper to the theory of variations of Hodge structures 
over $SL(N,\complex)$ bundles on $\Sigma$ which lies at the heart of W-gravity.
Moreover, the description of W-gravity in terms of an integrable hierarchy could provide through the duality
a new perspective on the r\^ole of integrable hierarchies in ${\cal N}=2$ gauge theories deformed by chiral operators. 

{\bf Acknowledgments}:
We thank  L.~F.~Alday and F.~Benini for useful comments on the manuscript and S.~Cecotti and 
N.~Nekrasov for stimulating discussions.

\end{document}